\input epsf

\documentstyle[amsfonts]{mn} 

\def\beq{\begin{equation}}
\def\eeq{\end{equation}}
\def\bey{\begin{eqnarray}}
\def\eey{\end{eqnarray}}
\def\beyn{\begin{eqnarray*}}
\def\eeyn{\end{eqnarray*}}

\def\d{{\mathrm{d}}}
\def\vecr{\bmath{r}}
\def\vecv{\bmath{v}}

\def\losvd{\textsc{losvd}}
\def\los{\textsc{los}}
\def\pvd{\textsc{pvd}}
\def\df{\textsc{df}}
\def\zvc{\textsc{zvc}}

\title{Probing a regular orbit with spectral dynamics}

\author[Y.\ Copin, H.S.\ Zhao, P.T.\ de Zeeuw]%
{Y.\ Copin$^1$ \thanks{E-mail: ycopin@obs.univ-lyon1.fr}, 
  H.S.\ Zhao$^2$, P.T.\ de Zeeuw$^2$ \\
  $^1$Centre de Recherche Astronomique de Lyon 
  \thanks{UMR CNRS 5574, \'Ecole normale sup\'erieure de Lyon - 
    Universit\'e Claude Bernard Lyon I}, 
  Observatoire de Lyon, 69561 Saint-Genis-Laval cedex, France \\
  $^2$Sterrewacht Leiden, Niels Bohrweg 2, 2333 CA, Leiden, The Netherlands
  }

\date{Accepted ........; Received ........; In original form .........}

\pagerange{\pageref{firstpage}--\pageref{lastpage}}

\pubyear{2000}

\begin{document}

\maketitle

\label{firstpage}


\begin{abstract}
We have extended the spectral dynamics formalism introduced by Binney
\& Spergel, and have implemented a semi-analytic method to
represent regular orbits in any potential, making full use of their
regularity. We use the spectral analysis code of Carpintero \& Aguilar
to determine the nature of an orbit (irregular, regular, resonant,
periodic) from a short-time numerical integration. If the orbit
is regular, we approximate it by a truncated Fourier time series of a
few tens of terms per coordinate. Switching to a description in
action-angle variables, this corresponds to a reconstruction of the
underlying invariant torus.  We then relate the uniform distribution
of a regular orbit on its torus to the non-uniform distribution in the
space of observables by a simple Jacobian transformation between the
two sets of coordinates. This allows us to compute, in a
cell-independent way, all the physical quantities needed in the study
of the orbit, including the density and in the line-of-sight velocity
distribution, with much increased accuracy. The resulting flexibility
in the determination of the orbital properties, and the drastic
reduction of storage space for the orbit library, provide a
significant improvement in the practical application of
Schwarzschild's orbit superposition method for constructing galaxy
models.  We test and apply our method to two-dimensional orbits in
elongated discs, and to the meridional motion in axisymmetric
potentials, and show that for a given accuracy, the spectral dynamics
formalism requires an order of magnitude fewer computations than the
more traditional approaches.
\end{abstract}



\section{Introduction}

In the construction of dynamical models for galaxies by
Schwarzschild's (1979) method, one tries to match the density
distribution and the photometric and kinematic observations with
weighted contributions of individual orbits (e.g., Rix et al.\ 1997;
Cretton et al.\ 1999).  Therefore, the method requires the precise
knowledge of the intrinsic properties of the orbits, such as the
density and the line-of-sight velocity distributions (\losvd's).
The traditional method for computing the orbital properties is to
integrate the equations of motion by numerical means. Then the spatial
density distribution of the orbit follows from calculating the
fraction of time spent in each cell of a grid in configuration space
after a sufficiently long integration period. Other properties follow
by considering similar grids in phase-space.  This method can be
visualized as simply marching along the orbit, and dropping balls in a
grid of buckets at regular time intervals.  It is conceptually similar
to a Monte-Carlo integration, and can be applied to any
orbit. However, this approach does not take advantage of the regular
behavior of many orbits in the potentials relevant for galaxies: it
simply treats the regular orbital motion as a structureless collection
of independent points in phase-space.

The motion in a regular orbit is quasi-periodic, and can be expressed
as a Fourier series expansion in action-angle variables. This series
expansion represents the underlying invariant orbital torus (e.g.,
Arnold 1989), and can be reconstructed from a normal orbit integration
(Binney \& Spergel 1982, 1984), or from generating functions (McGill
\& Binney 1990, Binney \& Kumar 1993, Kaasalainen \& Binney 1994,
Kaasalainen 1994). It is also possible to integrate the orbit directly
in the angle variables (Ratcliff, Chang \& Schwarzschild 1984),
although published applications have been restricted to two-dimensional
motion. Laskar (1993) introduced the so-called frequency map analysis
to study the nature of orbits.  This has been used to study orbits in
galactic potentials (Papaphilippou \& Laskar 1996, 1998; Valluri \&
Merritt 1998). Carpintero \& Aguilar (1998, hereafter CA98) developed
a fully automated technique to classify orbits based on the
commensurability of the peak frequencies in the Fourier spectra of the
orbit.

In this paper we extend this previous work as follows.  From a
standard numerical integration of the orbit over a few hundred orbital
periods, and after a detailed spectral analysis based on the method of
CA98, we obtain a semi-analytic expression for the orbit. This can be
considered as fitting the underlying orbital torus to the results of a
short numerical integration, and so obtaining an approximate
expression for the orbit valid for all time (i.e., fully phase-mixed).
It is then straightforward to derive all the required orbital
properties, including the density in configuration space and the
\losvd's, with high precision, and we describe in detail how to do
this.  While there are many papers on orbital tori reconstruction in
the literature, we are not aware of any published formalism for
projecting the tori to observable space to get the \losvd\ of an orbit.
Here we implement our method on two-dimensional integrable and
non-integrable potentials, and compare the results with traditional
straight numerical integrations.  We also show how to generalize
the results to axisymmetric three-dimensional potential.

This paper is organized as follows. After reviewing some basic results
of spectral dynamics in \S\ref{sec:introform}, we present the extended
formalism in \S\ref{sec:formalism}. In
\S\ref{sec:numeric} we test the formalism on a two-dimensional
separable (St\"ackel) potential, and then apply it to Binney's
logarithmic potential. We describe the application to the motion in
the meridional plane of an axisymmetric potential in
\S\ref{sec:3Daxi}, and summarize our conclusions in \S\ref{sec:concl}.


\section{Introduction to spectral dynamics}
\label{sec:introform}

For the convenience of the general reader, we here summarize some
results on the structure of phase-space, and introduce the terminology
and notations of action-angle description of orbits.


\subsection{Phase-space structure, action-angle variables 
  and base frequencies}
\label{sec:action}

In an $n$-dimensional potential (cases of interest are $n=2$ or 3), an
orbit can be characterized according to its $m$ integrals.  In a
time-independent potential, the energy is a conserved quantity and
is always an integral for an orbit, leading to $m \geq
1$.  When the orbital motion in phase-space (of dimension $2n$) is
constrained by the conservation of $m$ integrals of motion, the
orbital manifold on which the orbit evolves has dimension $2n-m$.
The orbit is said to be \emph{regular} if $m \geq n$, and then travels
on a manifold of dimension $\leq n$; otherwise ($m < n$), the orbit is
said to be \emph{irregular} (or \emph{chaotic}), and move in a region
of phase-space of dimension $>n$ (but still $\leq 2n-1$).

\begin{figure}
\epsfxsize=0.9\columnwidth
\centerline{\epsfbox{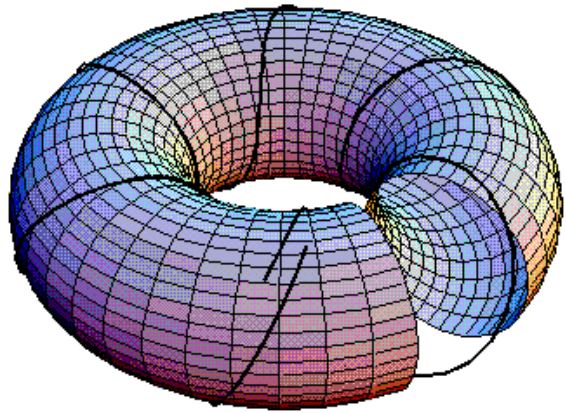}}
\caption{A regular open orbit spiraling on its invariant torus.}
\label{fig:torus}
\end{figure}

In the special case $m = n$, the orbital manifold of
a bound orbit is topologically equivalent to an $n$-torus, the
\emph{invariant torus}, embedded in phase-space (e.g., Arnold 1989;
see Fig.~\ref{fig:torus}), and the orbital motion is
\emph{quasi-periodic}.  Not being further constrained by additional
integrals of motion, the trajectory is \emph{dense} (or
\emph{ergodic}) on the torus, and the orbit is said to be
\emph{open}. On the other hand, if $m>n$, the orbit is further
constrained, and will not cover densely its torus even after infinite
time. This happens when there exist resonances between the fundamental
frequencies of two or more angle variables describing the orbit (cf.
following section).  Thus, we speak of \emph{non-resonant} ($m=n$) and
\emph{resonant} ($m>n$) orbital tori; only the former form a set of
non-zero measure in phase-space. If the orbit is \emph{fully
resonant}, i.e., $m=2n-1$, it closes on itself only after a finite
number of turns around its torus: such an orbit is \emph{periodic}.

A regular orbit is confined to its invariant $n$-torus, so we can look
for canonical coordinates $(\bmath{Q},\bmath{P})$ adapted to the torus
geometry, such that the momenta $\bmath{P}$ are constant on the torus,
and their conjugate variables $\bmath{Q}$ form a natural coordinate
system for the region of phase-space occupied by the tori. Since these
coordinates are canonical, Hamilton's equations state:
\beq
\dot{\bmath{Q}} = \frac{\partial H}{\partial \bmath{P}}, \qquad 
\dot{\bmath{P}} = -\frac{\partial H}{\partial \bmath{Q}},
\eeq
where $H$ is the Hamiltonian of the orbit. Since we want $\bmath{P}$
to be constant, this implies that $H = H(\bmath{P})$, i.e., that $H$
is cyclic in $\bmath{Q}$, and thus that $\partial H / \partial
\bmath{P} = \mathrm{cte} = \bomega$. We can therefore integrate the
remaining equation of motion straightforwardly: $\bmath{Q}(t) =
\bomega\,t + \bmath{Q}(0)$.

Choosing an appropriate scaling such that $\bmath{Q}$ is
$2\pi$-periodic, we obtain the so-called \emph{action-angle
  variables}: the angles $\bvarphi \equiv \bmath{Q}$ are natural
coordinates for the invariant torus, and allow one to distinguish
individual points on it; the actions $\bmath{\mathcal J} \equiv \bmath{P}$
label this torus with respect to the tori of other orbits. They can be
seen as related to the different radii of the tori.

If the frequencies (actually pulsations) $\omega_i$ ($i=1,\ldots,n$)
are mutually incommensurable, they are said to be the $n_{\mathrm{BF}}
= n$ \emph{base frequencies} of the orbit; the incommensurability
ensures that the orbit covers densely its torus and therefore is open.
If there exist a linear commensurability between two or more
$\omega_i$, then the orbit is further constrained on the torus, and
the number of incommensurable base frequencies is reduced to
$n_{\mathrm{BF}} < n$. In the extreme case, if the orbit is closed,
only one base frequency remains. In the other extreme case, the base
frequencies are not well-defined for irregular orbits, since these are
not constrained to a $n$-torus; however, according to the practical
definition of the base frequencies, they will be characterized by
$n_{\mathrm{BF}} > n$ (CA98).


\subsection{Description of an orbit in angle variables}

We now turn to the two-dimensional case, $n=2$, and restrict ourselves
to \emph{regular open orbits}, which leads to $n_{\mathrm{BF}} = n =
2$. The other regular orbits (with $n_{\mathrm{BF}} < 2$) can be
considered as a degenerate subset of this case, in the sense that the
resonances lead to invariant tori of zero measure in phase space.

A regular open orbit in a two-dimensional potential is characterized
by its two base frequencies $\bomega \cor 2\pi\bnu$. The angle
variables $\bvarphi \in \left[0,2\pi\right[$ are then given by
\beq
\label{eq:defphi}
\bvarphi(t) = \bomega t,
\eeq
where we have chosen the zero of time such that $\bvarphi(0) =
\bmath{0}$.  We can then write the position $\vecr\equiv (x,y)$
and the velocity $\vecv \cor \dot{\vecr}$ at time $t$ as
\beq
\vecr(t) \equiv \vecr(\bvarphi(t)) \qquad\mbox{and}\qquad
        \vecv(t) \equiv \vecv(\bvarphi(t)).
\eeq
The quasi-periodicity of the orbit allows us to write the orbital
motion in a Fourier form (Binney \& Spergel, 1982):
\beq
\label{eq:quasi}
\left\{ \begin{array}{rcl}
        x(t) &=& \sum\limits_{l,m} X_{(l,m)} 
                \cos(\omega_{(l,m)} t + \chi_{(l,m)}), \\
        y(t) &=& \sum\limits_{l,m} Y_{(l,m)} 
                \cos(\omega_{(l,m)} t + \psi_{(l,m)}),
        \end{array}\right.
\eeq
where ${(l,m)}\subset {\mathbb Z}^2$ are pairs of integers,
$\chi_{(l,m)}$ and $\psi_{(l,m)}$ are constant phases,
and the frequencies $\omega_{(l,m)}$ are linear combinations of
the base frequencies:
\beq
\omega_{(l,m)} \equiv l\omega_1 + m\omega_2  \cor \blambda\cdot\bomega,
\eeq
and we have used the shorthand $\blambda \equiv (l,m)$.
Alternatively we can write 
\beq
x(t) = \sum_{\blambda} X_\blambda 
         \cos(\blambda\cdot\bomega\,t + \chi_\blambda),
\eeq
for the position and
\beq
v_x(t) = - \sum_{\blambda}  \blambda\cdot\bomega \, X_\blambda
         \sin(\blambda\cdot\bomega\,t + \chi_\blambda)
\eeq
for the velocity, and similar expressions for $y(t)$ and $v_y(t)$. 

Switching from time $t$ to angle variables $\bvarphi$ by means of
eq.~(\ref{eq:defphi}), we obtain the torus expression of the motion
(in what we will refer to as the $\varphi$-space hereafter), in the
sense that angle variables are natural coordinates for the invariant
torus on which the regular orbit evolves:
\beq
\label{eq:xphi}
x(\bvarphi) = \sum_{\blambda} X_\blambda 
        \cos(\blambda\cdot\bvarphi + \chi_\blambda),
\eeq
\beq
\label{eq:vphi}
v_x(\bvarphi) = - \sum_{\blambda} X_\blambda\, \blambda\cdot\bomega
        \sin(\blambda\cdot\bvarphi + \chi_\blambda), 
\eeq
and similar expressions for $y(\bvarphi)$ and $v_y(\bvarphi)$.


\section{Extended formalism}
\label{sec:formalism}


\subsection{$\bvarphi$-inversion}
\label{sec:inversion}

The first step for calculating the density $\rho$ at a point
$\vecr_0=(x_0,y_0)$ in configuration space is to find the
corresponding actions and angles at that point.  However, while
$\vecr(\bvarphi)$ and $\dot{\vecr}(\bvarphi)$ (cf.
eqs~(\ref{eq:xphi}) and~(\ref{eq:vphi})) are single-valued
functions, the inverse functions $\bvarphi(\vecr)$ and
$\bvarphi(\dot{\vecr})$ are generally multi-valued.  For instance, a
given position $\vecr_0$ on the orbit corresponds in general to a
still-to-be-defined number $N$ of solutions $\bvarphi^{(i)}_0$ such
that:
\beq
\label{eq:phiofr}
\vecr(\bvarphi^{(i)}_0) = \vecr_0, \qquad i=1,\ldots,N.
\eeq
The total number $N$ of solutions is finite (see, e.g.,
Fig.~\ref{fig:xyofphi}).  We will refer to the $N$ different
single-valued functions involved in the definition of $\bvarphi$ by
$\bvarphi^{(i)}$, such that $\vecr(\bvarphi^{(i)})$ can be inverted.

The non-uniqueness of the solution can be understood, in the case of
$\bvarphi(\vecr)$ for instance, by the fact that an orbit usually has
a finite number of ways to reach a given point $\vecr_0$, either from
a different part of the trajectory, or along the same part of
trajectory but in different directions in a case of an orbit without a
given sense of propagation (such as a box orbit). Accordingly, for a
given orbit we can predict the general number of solutions $N$
(excepting the degenerate cases) of eq.~(\ref{eq:phiofr}).
Specifically, $N=2$ for an open two-dimensional tube orbit, since the
orbit can usually access a given point along two paths with a given
direction of propagation imposed by the fixed sense of rotation.
$N=1$ on the boundaries of the orbit. For a flat box orbit each of the
two paths through a given point can be traveled in both directions
since there is no fixed sense of rotation, so that $N=4$. On the outer
boundaries of the box, degeneracy results in $N=2$, or even $N=1$ in
the corners of the orbit, which are accessible in only one way (the
one perpendicular to the zero-velocity curve; see Ollongren 1962).
Each time a resonance brings two distinct parts of the orbit on top of
each other, it is easy to see that the number of solutions in the zone
of overlap will be doubled. For instance, we expect $N=8$ in the zone
of overlap of an open $3:2$ (`fish') orbit (Miralda-Escud\'e
\& Schwarzschild 1989; CA98).


\subsection{Orbital density}
\label{sec:density}

In configuration space, we can define the orbital density $\rho$ at
point $\vecr_0$ by writing the mass of the orbit enclosed in the
element of volume $d\tau = dx\,dy$ around $\vecr_0$ as
\beq
\label{eq:dmtau}
\d m \cor \rho(\vecr_0)\,\d\tau.
\eeq
Equivalently, in $\varphi$-space, we can define the density
$\rho_\varphi$ at a point $\bvarphi_0$ such that the mass enclosed in
the element of volume $d\phi = d\varphi_1\,d\varphi_2$ around
$\bvarphi_0$ is:
\beq
\label{eq:dmphi}
\d m' \cor \rho_\varphi(\bvarphi_0)\,\d\phi.
\eeq
If we choose to normalize the total mass of the orbit to~1, then 
$\int\!\d m = \int\!\d m' = 1$.

An open orbit, i.e., an orbit with two incommensurable
base frequencies, will eventually fill its torus uniformly, so that
the density $\rho_\varphi$ in $\varphi$-space is constant (`time
averages theorem', BT87).  Mass normalization then gives:
\beq
\label{eq:normdensphi}
\rho_\varphi = \frac{1}{\int\!\d\phi} = \frac{1}{(2\pi)^2}.
\eeq
If the orbit is closed, i.e., the orbit has only one base frequency,
all the other frequencies in eq.~(\ref{eq:quasi}) being commensurable
with it, then the time averages theorem does not apply, since the
orbit is confined to a closed spiral on its torus by a new integral of
motion.

Using the $N$ single-valued functions $\bvarphi^{(i)}$ giving the
solutions $\bvarphi^{(i)}_0$ of eq.~(\ref{eq:phiofr}), we
can relate the element of volume $\d\tau$ around the point $\vecr_0$
to the elements of volume $\d\phi^{(i)}$ around $\bvarphi^{(i)}_0$:
\beq
\d\phi^{(i)} = \left| \frac{\partial \bvarphi^{(i)}}
        {\partial \vecr}\right|_{\vecr_0}\!\!\d\tau
        = \frac{\d\tau}{{J_{\vecr}^{(i)}}},
\eeq
where we have defined the Jacobian $J_{\vecr}^{(i)}$ as in 
eq.~(\ref{eq:jacB}):
\beq
\label{eq:jacr}
J_{\vecr}^{(i)} \cor \left| \frac{\partial \vecr}
        {\partial \bvarphi}\right|_{\bvarphi^{(i)}_0}. 
\eeq
We can then link the density $\rho$ to the constant density
$\rho_\varphi$ given by eq.~(\ref{eq:normdensphi}) by properly
relating eqs~(\ref{eq:dmtau}) and~(\ref{eq:dmphi}):
\beq
\d m \equiv \sum_{i=1}^{N} \d m^{(i)},
\eeq
with $\d m^{(i)} \cor \rho_\varphi(\bvarphi^{(i)}_0)\,\d\phi^{(i)}$, 
since the mass lying in $\d\tau$ around a given point
is the sum of the masses lying in the various elements $\d\phi^{(i)}$ 
contributing to $\d\tau$. This gives:
\beq
\d m = \frac{\d\tau}{(2\pi)^2}\,\sum_{i=1}^{N} \frac{1}{{J_{\vecr}^{(i)}}},
\eeq
which, by comparison with eq.~(\ref{eq:dmtau}), leads to the final expression:
\beq
\label{eq:rho}
\rho(\vecr_0) = \frac{1}{(2\pi)^2}\,
        \sum_{i=1}^{N}\frac{1}{J_{\vecr}^{(i)}},
\eeq
the Jacobians $J_{\vecr}^{(i)}$ being computed at the $N$ solutions
$\bvarphi^{(i)}_0$ of eq.~(\ref{eq:phiofr}). The various
$J_{\vecr}^{(i)}$ do not have to be the same, since the functions
$\bvarphi^{(i)}$ are different.


\subsection{Boundary of the orbit}
\label{sec:boundary}

The boundary $\mathcal B$ of an orbit can be considered as the
location of the points leading to the degeneracy of solutions of
eq.~(\ref{eq:phiofr}). At a point $\vecr_{\mathcal B}$ of
$\mathcal B$, we can say that if $\bvarphi_{\mathcal B}$ is a
solution of $\vecr(\bvarphi) = \vecr_{\mathcal B}$, then
$\vecr(\bvarphi_{\mathcal B} + \delta\bvarphi)$ is also a solution to
the second-order in $\|\delta\bvarphi\|$; we have:
\beyn
\vecr(\bvarphi_{\mathcal B} + \delta\bvarphi) 
        & \simeq & \vecr(\bvarphi_{\mathcal B}) + 
        \left.\frac{\partial \vecr}{\partial \bvarphi}
                \right|_{\bvarphi_{\mathcal B}}\!\!\times\delta\bvarphi
        \qquad \|\delta\bvarphi\| \to 0 \\
        & = & \vecr_{\mathcal B} + O(\|\delta\bvarphi\|^2)
\eeyn
if and only if 
\beq
\label{eq:jacB}
J_{\vecr}(\bvarphi_{\mathcal B}) \cor \left|\frac{\partial \vecr}
        {\partial \bvarphi}\right|_{\bvarphi_{\mathcal B}} = 0. 
\eeq
We then reach another equivalent definition of the boundary
$\mathcal B$, as the location of the points where at least one
Jacobian $J_{\vecr}$ vanishes (and possibly more, since there is a
degeneracy of the solutions at this point).

At the boundary, the orbit is divergent in the density (cf.
eq.~(\ref{eq:rho})), but this divergence is integrable since the total
mass of the orbit is finite.


\subsection{Orbital distribution function}
\label{sec:orbital_df}

The above considerations, related to equations analogous to
(\ref{eq:xphi}) and (\ref{eq:vphi}), directly lead to an expression of
the orbital distribution function~(\df) $f(\vecr_0,\vecv_0)$, by
considering $\left[(2\pi)^2\,J_{\vecr}^{(i)}\right]^{-1}$ as the
contribution to the local density $\rho$ from solution $i$ among $N$:
\beq
\label{eq:dfr}
f(\vecr_0,\vecv_0) =
        \frac{1}{(2\pi)^2} \sum_{i=1}^N \frac{1}{J_{\vecr}^{(i)}} \times
        \delta\left(\vecv_0 - \vecv(\bvarphi^{(i)})\right), 
\eeq
where $\bvarphi^{(i)}, i=1,\ldots,N$ are the solutions to
eq.~(\ref{eq:phiofr}).  Given the relation~(\ref{eq:rho}), this
expression is consistent with the basic relation:
\beq
\rho(\vecr) \cor \int f(\vecr,\vecv)\,\d\vecv,
\eeq
where the velocity is an indirect function of the position.
Eq.~(\ref{eq:dfr}) gives the~\df\ at any position $\vecr_0$.

There are several variations for expressing the phase space density,
depending whether we are primarily interested in the distribution of
the position or of the velocity.  For generalization, let us consider
a vector $\bmath{\kappa}$ from a half-subspace (of dimension $n=2$) of
phase-space (of dimension $2n = 4$), and $\bmath{\iota}$ from the
complementary half-subspace. The previous situation corresponded to
$\bmath{\kappa} = \vecr \equiv (x,y)$ and $\bmath{\iota} = \vecv
\equiv (v_x,v_y)$, but we could have, as we shall see in
\S\ref{sec:losvd}, $\bmath{\kappa} \equiv
(x_{\mathrm{sky}},v_{\mathrm{los}})$ and $\bmath{\iota} \equiv
(x_{\mathrm{los}},v_{\mathrm{sky}})$. We now look for an expression of
the~\df\ $f(\bmath{\kappa}_0,\bmath{\iota}_0)$ from $\bmath{\kappa}$.

In the same way as in the special case $\bmath{\kappa} = \vecr$
described in \S\ref{sec:density}, one can show that:
\beq
\label{eq:dfgen}
f(\bmath{\kappa}_0,\bmath{\iota}_0) =
        \frac{1}{(2\pi)^2} \sum_{i=1}^M \frac{1}{J_{\bmath{\kappa}}^{(i)}} 
        \times
        \delta\left(\bmath{\iota}_0 - \bmath{\iota}(\bvarphi^{(i)})\right),
\eeq
where this time, we have defined $J_{\bmath{\kappa}}^{(i)}$ as:
\beq
\label{eq:jac}
J_{\bmath{\kappa}}^{(i)} \cor \left| \frac{\partial \bmath{\kappa}}
        {\partial \bvarphi}\right|_{\bvarphi_0^{(i)}}, 
\eeq
and with $\bvarphi_0^{(i)}$ being the multiple solutions of
the equation:
\beq
\label{eq:phiofkappa}
\bmath{\kappa}(\bvarphi^{(i)}_0) = \bmath{\kappa}_0, \qquad i=1,\ldots,M.
\eeq
The number $M$ of solutions does not have to be the same as the number
$N$ of solutions to eq.~(\ref{eq:phiofr}).


\subsection{Line-of-sight velocity distribution}
\label{sec:losvd}

Eq.~(\ref{eq:dfr}) provides the \df\ $f(\vecr,\vecv)$ at any point of
phase-space accessible to the orbit, and hence allows computation of
all the orbital dynamical quantities, directly related to the~\df, in
particular the line-of-sight velocity distribution~(\losvd).

Once we have chosen a line-of-sight (\los), associated with
coordinates $(x',y')$ ($y'$ along the~\los\ and $x'$ being in the
plane of the sky), we can express the \losvd\ 
${\mathrm{VP}}_{x'}(v_{y'})$ (the \emph{velocity profile}) by the
usual formula:
\beq
{\mathrm{VP}}_{x'}(v_{y'}) \cor 
        \int_{\mathrm{los}}\!\d y' \int \d v_{x'} f(\bmath{r'},\bmath{v'}).
\eeq
We see that, for this purpose, we need an expression for the~\df\ as a
function of $\bmath{\zeta} \equiv (x',v_{y'})$. According to
eq.~(\ref{eq:dfgen}), we can write:
\beq
\label{eq:losvd}
{\mathrm{VP}}_{x'_0}(v_0) = \frac{1}{(2\pi)^2}\,
        \sum_{i=1}^{M}\frac{1}{J_{\bmath{\zeta}}^{(i)}},
\eeq
where the Jacobians $J_{\bmath{\zeta}}^{(i)}$ are to be computed 
at the $M$ solutions of the equation:
\beq
\label{eq:phioflosvd}
\bmath{\zeta}(\bvarphi^{(i)}_0) = \bmath{\zeta}_0 \qquad \iff \quad
\left\{ \begin{array}{rcl}
        x'(\bvarphi^{(i)}) &=& x'_0, \\
        v_{y'}(\bvarphi^{(i)}) &=& v_0.
\end{array}\right.
\eeq
We can also write down the surface brightness $\mu(x')$ as:
\beyn
\mu(x') & \cor & \int_{\mathrm{los}}\!\d y' \int \d\bmath{v'}
        f(\bmath{r'},\bmath{v'}) \\
        & = & \int_{\mathrm{los}}\! \rho(\bmath{r'})\,\d y'
        = \int {\mathrm{VP}}_{x'}(v)\,\d v,
\eeyn 
which gives two equivalent ways to compute these quantities, either
from the density, or from the \losvd.


\subsection{Actions}

The formalism we use here, based on an expression of the dynamical
quantities of the orbit in terms of the angle variables $\bvarphi$, is
well-suited for the computation of the actions.  An open
two-dimensional regular orbit is characterized by exactly two
integrals of motion, which can be for instance the two actions
${\mathcal J}_1$ and ${\mathcal J}_2$, that can be defined by (BT87):
\beq
\label{eq:def_actions}
{\mathcal J}_i \cor \frac{1}{2\pi} 
        \!\!\oint\limits_{\scriptstyle \varphi_i \in\left[0,2\pi\right[} 
        \!\!\!\!\vecv \cdot \d\vecr, \qquad i=1,2.
\eeq
Using relations~(\ref{eq:xphi}) and~(\ref{eq:vphi}), we obtain after
some algebra:
\beq
\label{eq:action}
{\mathcal J}_{1 \choose 2} = \frac{1}{2}\sum_{\blambda}\left[ 
  (X_\blambda^2 + Y_\blambda^2) {l \choose m}\blambda\cdot\bomega\right].
\eeq
which is very similar to eq.~(16) in Binney \& Spergel (1984). As
explained by these authors, one is free to define the `real' (useful)
actions ${\mathcal J}_r$ and ${\mathcal J}_a$ as any linear
combination of the previous quantities, so that it matches any natural
requirement such as ${\mathcal J}_r \equiv 0$ for closed long-axis
orbits and ${\mathcal J}_a \equiv 0$ for closed loop orbits (BT87). In our
case, with the CA98 extraction procedure, we found that the linear
transformation to apply was (cf. Fig.~\ref{fig:action}):
\beq
\left\{ \begin{array}{rcl}
        {\mathcal J}_r & = & {\mathcal J}_1 \\
        {\mathcal J}_a & = & {\mathcal J}_2
        \end{array}\right. \quad \mbox{for box orbits,}
\eeq
and
\beq
\left\{ \begin{array}{rcl}
        {\mathcal J}_r & = & 2{\mathcal J}_2 \\
        {\mathcal J}_a & = & {\mathcal J}_1 - {\mathcal J}_2
        \end{array}\right. \quad \mbox{for loop orbits.}
\eeq
This choice guarantees a continous action space 
(cf. Binney \& Spergel 1984; de Zeeuw 1985).  


\section{2D numerical implementation}
\label{sec:numeric}

We have numerically implemented the formalism described above, and in
\S\ref{sec:numstack} describe tests on orbits in a two-dimensional
separable potential for which all relevant quantities are known
analytically.  We then consider a non-integrable logarithmic potential
in \S\ref{sec:logpot}.


\subsection{Sridhar \& Touma potential}
\label{sec:numstack}

We first study an orbit in a two-dimensional separable potential,
described by Sridhar \& Touma (1997; see Appendix~\ref{app:stackel}).
This potential is of St\"ackel form in parabolic coordinates, and
allows a complete analytic study of the orbits (in particular
computation of the boundaries and expression of the surface density),
which are all regular, and usually denoted as banana orbits.

The illustrations in this section are made for a slightly cusped
potential ($\alpha=0.5$, with notations of
Appendix~\ref{app:stackel}), and a test-orbit with energy $E=2.39$ and
second integral $I_2=0.38$, integrated on 4096~points over 123~periods
with a 7/8$^{\mathrm{th}}$ order Runge-Kutta integrator
(Fig.~\ref{fig:reconst}).

\subsubsection{Spectral analysis and time-dependent expressions}
\label{sec:numspec}

\begin{figure}
\epsfxsize=0.9\columnwidth
\centerline{\epsfbox{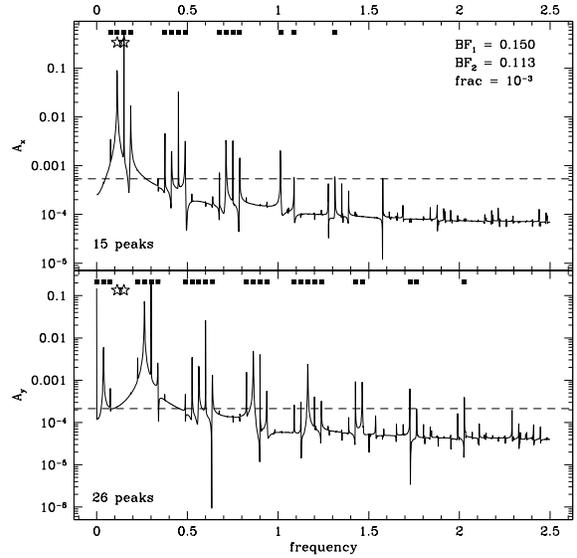}}
\caption{Fourier spectra (amplitude vs.\ frequency) for the test orbit. 
  The selected frequencies for the decomposition (\emph{squares}) are
  those of amplitude $A_i$ larger than $F \max[A_i]$ in each
  coordinate, indicated by the \emph{long-dashed} lines, with
  $F=10^{-3}$ (giving 15~peaks in $x$ and 26~in $y$).  The two base
  frequencies are marked by \emph{stars}.}
\label{fig:fourier}
\end{figure}

The Fourier spectra $\hat{x}(\nu)$ and $\hat{y}(\nu)$ of the time
series of the coordinates $x(t)$ and $y(t)$ of a quasiperiodic orbit,
obtained in the usual way by numerical integration, consist of
discrete lines whose frequencies are linear combinations of the base
frequencies. We modified an algorithm provided by CA98 for the
extraction of the base frequencies, in order to give all the
quantities defining the Fourier series (\ref{eq:quasi}).  The CA98
method computes the Fourier transform of the coordinates of an
integrated orbit, identifies the peaks in the Fourier spectra and
extracts the corresponding frequencies. It looks for the base
frequencies, and linearly decomposes the selected frequencies over
these BFs (see Fig.~\ref{fig:fourier}).  The output then consists of:
the base frequencies $\bomega$, and for each peak frequency in each
coordinate, the linear decomposition over the base frequencies (e.g.,
$\blambda\equiv(l,m)$), and the associated amplitude ($X_\blambda$)
and phase ($\chi_\blambda$).

\begin{figure}
\epsfxsize=0.9\columnwidth
\centerline{\epsfbox{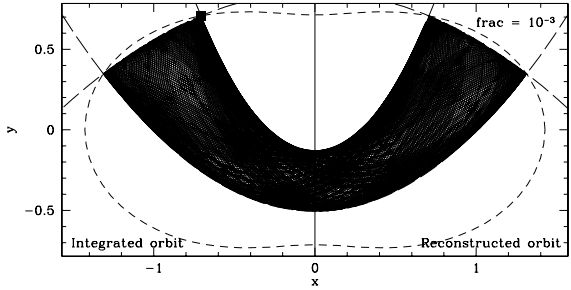}}
\caption{Comparison between time expression from numerical integration
  (\emph{left}) and spectral expansion (\emph{right}) of an orbit in
  the Shridhar \& Touma potential. The spectral time-dependent
  expansion (\ref{eq:quasi}) was obtained with $F=10^{-3}$ (resulting
  in 15~frequencies in $x$ and 26~in $y$) at the same time-steps as
  the integrated orbit. The \emph{short-dashed} line is the
  Zero-Velocity Curve (\zvc), and the \emph{long-dashed} curves
  indicate the boundaries of the orbit calculated by using the
  separable nature of the potential (cf. Appendix~\ref{app:stackel}).
  The orbit was launched from the \zvc\ (\emph{square}).  The 
  boundaries of the reconstructed orbit are barely distinguishable
  from the exact boundaries, indicating the level of precision of the
  expansion.}
\label{fig:reconst}
\end{figure}

The CA98 algorithm provides a finite number of terms in the
quasi-periodic expansion (\ref{eq:quasi}).  The selection is made by
keeping, in each coordinate, only frequencies with amplitude greater
than a fraction $F$ of the greatest one (typically $F =
10^{-2}$--$10^{-3}$, depending on the required accuracy), giving a
finite number of terms in each direction (typically between~10
and~20). The numerical approximations obtained by the truncation of
the series coincide with the actual expansion (\ref{eq:quasi}) up to
the required accuracy (Fig.~\ref{fig:reconst}). 

\subsubsection{$\varphi$-expressions}

The time-dependent expansion (\ref{eq:quasi}) provides a semi-analytic
expression for a regular orbit. It can be useful for storage purposes,
or for extrapolating orbital motion to times much longer than the
actual integration time (e.g., in order to reduce the numerical noise
of a grid-based density computation). However, time is not a good
variable for a regular orbit that will eventually fill up completely
its invariant torus, and the angle variables $\bvarphi$ should be used
instead. This leads to the expressions (\ref{eq:xphi}) and
(\ref{eq:vphi}) for $x(\bvarphi)$ and $y(\bvarphi)$, respectively.

\begin{figure}
\epsfxsize=0.9\columnwidth
\centerline{\epsfbox{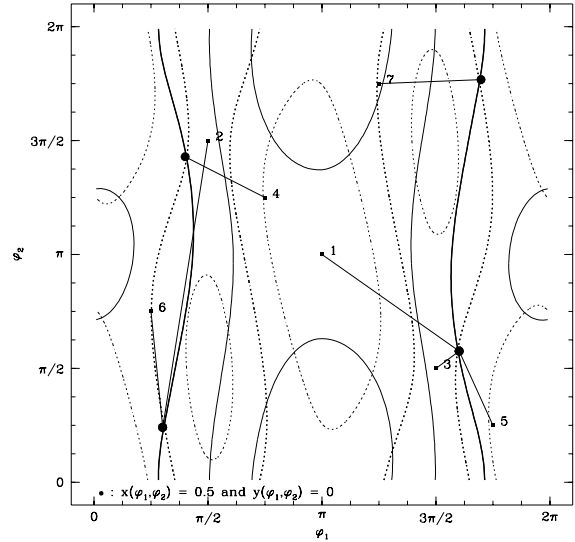}}
\caption{Contours of $x(\bvarphi)$ (\emph{solid} lines) and $y(\bvarphi)$ 
  (\emph{dotted} lines), for the test orbit reconstructed with $F=10^{-3}$.
  The levels $x=x_0=0.5$ and $y=y_0=0$ (\emph{thick} lines) intersect
  at the $N=4$ values of $\bvarphi(\vecr_0)$ (\emph{dots}).
  Successive initial conditions for the root-finding routine
  (\emph{squares}) lead to different solutions (\emph{straight}
  lines), until it finds the expected number of solutions.}
\label{fig:xyofphi}
\end{figure}

The problem is then to recover the $N$ different solutions of
eq.~(\ref{eq:phiofr}) for a given position $\vecr_0$. First, we have
to find the expected number $N$ of solutions.  This is inferred from
the orbit classification, according to the criteria described in
\S\ref{sec:inversion}. We then use standard root-finding routines (in
our case, the \textsc{nag} routine \textsc{c05pbf}) from different
initial conditions until we have found the $N$ solutions, or until a
maximum number of trials, indicating a degenerate case with fewer
solutions (see Fig.~\ref{fig:xyofphi}).  Interestingly, there is
point-symmetry in the solutions.  For example, for the Sridhar \&
Touma banana orbit in Fig.~\ref{fig:xyofphi}, the center of the
symmetry is at $\bvarphi = (\pi,\pi)$.  In fact this is true for
\emph{any} box orbit in a general non-rotating potential as well.
This is because we can always choose the corner of a box orbit as the
initial conditions, i.e., $\vecv = \bvarphi = \bmath{0}$ at $t=0$.
Then for any box described by $\vecr(t)$, we can generate the new
orbit described by $\vecr(-t)$ by reversing the arrow of the time, and
the new orbit simply \emph{repeats} the old one with $\vecr(-t) =
\vecr(t)$: reversing has no effect on the orbit because the initial
velocity is zero.  This time-symmetry means $\bvarphi\,[2\pi]$ and
$-\bvarphi\,[2\pi]$ describe the same point in the configuration
space, where the modulus brings the phase angle to the default
interval $[0,2\pi]$.  Thus the solutions come in pairs at $\bvarphi$
and $2\pi-\bvarphi$ with point-symmetry around $(\pi,\pi)$.  The above
arguments do not apply to loop orbits because there is no point in a
loop where the velocity is zero\footnote{Reversing the arrow of time
  from a initial point at say, $x=0,\,y=1$ of a loop, will lead to a
  counter-loop with identical shape indeed, but for which $\bvarphi$
  and $2\pi-\bvarphi$ correspond to two different points in
  configuration space, with $(x(-t),y(-t)) = (-x(t),y(t))$.}.

\subsubsection{Density maps and velocity profiles}
\label{sec:densmaps}

\begin{figure}
\epsfxsize=0.9\columnwidth
\centerline{\epsfbox{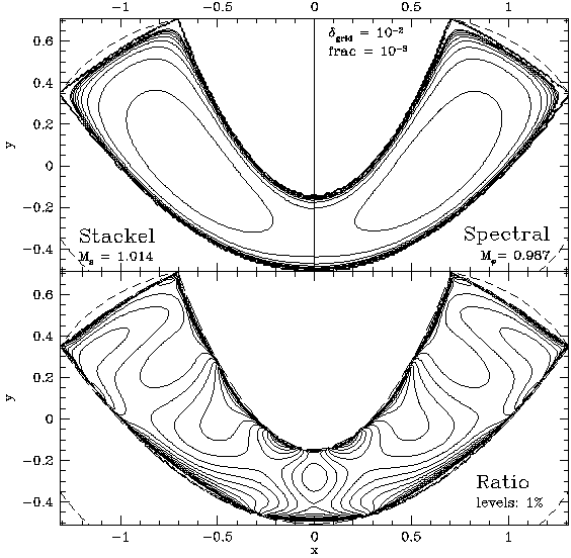}}
\caption{\emph{Upper panel}: Orbital density distribution in the Shridhar 
  \& Touma potential computed from spectral theory (\emph{right})
  compared with the exact expression (\emph{left}).  The spectral
  density~(\ref{eq:rho}) was obtained with $F=10^{-3}$ on the same
  grid as the St\"ackel density (step
  $\delta_{\mathrm{grid}}=10^{-2}$). The density levels range from
  0~to~2 in 11~steps. The total mass of the orbit is normalized to 1
  by definition in each case.  \emph{Lower panel}: Ratio between
  spectral density and St\"ackel density, in the case
  $\mathrm{frac}=10^{-3}$. The contour levels range linearly from
  $0.9$ to $1.1$ in steps of $0.01$ (\emph{thick} line $\equiv 1$). The
  increased discrepancy close to the boundary of the orbit is due to
  the divergence of the density and its associated numerical
  imprecision.}
\label{fig:compare}
\end{figure}

Eq.~(\ref{eq:rho}) allows the computation of the orbital density at
any point $\vecr_0$ once we have determined the $N$ solutions of
eq.~(\ref{eq:phiofr}) by the appropriate root-finding routine. Since
the test orbit was integrated in a St\"ackel potential, the derived
density distribution can be compared to the analytic expression
available for this case (cf. Appendix~\ref{app:stackel}).
Fig.~\ref{fig:compare} shows contours of the two densities side by
side with the same levels. The spectral density was obtained from an
expansion truncated in amplitude with $F=10^{-3}$, giving 15~terms in
$x$ and 26~in $y$. The indicated masses, which should be unity by
construction in both cases, are computed from a numerical integration
over the density grid; it can be therefore used as a rough estimate 
of the precision of the density computation.

As expected, the more terms are kept in the series (\ref{eq:quasi}),
the more accurate the representation of the orbit
(Fig.~\ref{fig:compare}). With $F=10^{-2}$, the orbit is represented
by an expansion with 4 terms in $x$ and 12 terms in $y$, and the
density is recovered to $\sim 20\%$. When $F$ is set to $10^{-3}$
(Fig.~\ref{fig:compare}), the expansion contains 15~terms in $x$ and
26~in $y$, and the density is accurate to $\sim 3\%$ in the inner
parts of the orbit, and degrades very close to the boundaries of the
orbit due to the divergence of the density there. This precision in
density can be compared with the usual computation of the density,
directly during integration (see \S\ref{sec:CPU}).

\begin{figure}
\epsfxsize=0.9\linewidth
\centerline{\epsfbox{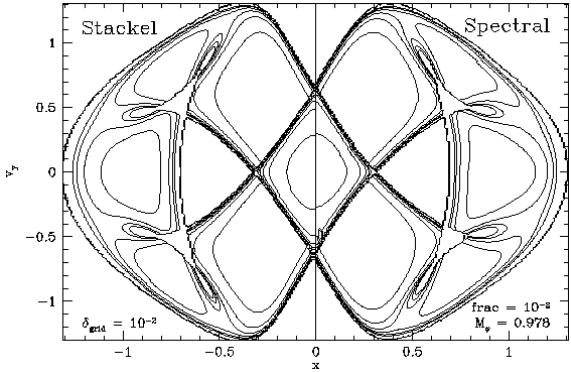}}
\caption{Position-velocity diagram (\pvd), 
  i.e., density in the plane $(x,\dot{y})$, for the test-orbit,
  computed from St\"ackel expression (\emph{left}) and from spectral
  theory with $F=10^{-3}$ (\emph{right}). The contours range linearly
  from 0~to~$0.4$ in 9~steps.  Since the orbit is a banana orbit,
  hence a boxlet orbit without a definite sense of rotation, the
  \pvd{} is symmetric with respect to $\dot{y}=0$.}
\label{fig:compare_losvd}
\end{figure}

Using eq.~(\ref{eq:losvd}), we can also compute the velocity profile
along any line-of-sight, and gather in a position-velocity diagram
(\pvd) all the velocity profiles corresponding to parallel~\los. This
is the projected density of the phase-space orbit on the plane
$(x',v_{y'})$ (where $x'$ and $y'$ are the coordinates running
perpendicular and parallel to the~\los), in the same way as the
`density' corresponds to the projected density of the phase-space
orbit on the plane $(x,y)$. A cut of this \pvd{} along a given $x'_0$
would give the \losvd\ along a line-of-sight running parallel to the
$y'$-axis and going through $x'_0$.  Fig.~\ref{fig:compare_losvd}
presents side by side the \pvd{} of the test-orbit for a~\los\ 
parallel to the $y$-axis, computed from the St\"ackel formalism
(\emph{left}) and from the spectral theory (\emph{right}), using the
same spectral expansion as before.  Precision of the computation is as
expected ($\sim 10\%$, see \S\ref{sec:dmvp}).


\subsection{Logarithmic potential}
\label{sec:logpot}

We now consider a non-integrable potential of astronomical interest,
the logarithmic potential (BT87, p.126):
\beq
\label{eq:logpot}
\Phi(x,y) \cor \frac{1}{2}v_0^2\, 
          \ln\left(R_c^2 + x^2 + \frac{y^2}{q^2}\right),
\eeq
with $v_0$ the circular speed at large radius, $R_c$ the core radius,
and $q$ the axial ratio of the equipotentials. Positivity of the
associated density requires $1/\sqrt{2}\leq q\leq 1$. This potential
admits two major families of orbits (BT87), the box orbits and the
flat-tube or loop orbits, accessible from different initial
conditions. For the illustrations, we consider the case $v_0\equiv
R_c=1$ and $q=0.9$.

\subsubsection{Density maps and velocity profiles}
\label{sec:dmvp}

\begin{figure}
\epsfxsize=0.9\columnwidth
\centerline{\epsfbox{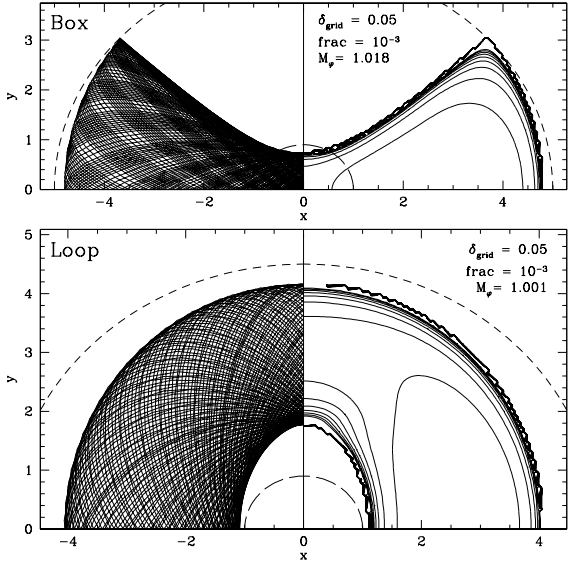}}
\caption{\emph{Upper panel}: Integrated box orbit (\emph{left}) 
  and spectral density (\emph{right}) computed from an expansion
  truncated at $F=10^{-3}$ on a grid of step
  $\delta_{\mathrm{grid}}=10^{-2}$.  The contours range from 0 to
  $0.1$ in 11~steps.  The \emph{short-dashed} line is the \zvc, and
  the \emph{long-dashed} line corresponds to the core of the
  potential.  \emph{Lower panel}: Same as previous, but for the
  $y>0$-side of the loop orbit.  The contours range from 0~to
  $5.10^{-2}$ in 11~steps.}
\label{fig:boxloopdens}
\end{figure}

Two orbits of the same energy $E=1.629$ were integrated: a box orbit
over 128~periods, and a loop orbit over 167~periods
(cf. Fig.~\ref{fig:boxloopdens}). This energy gives a \zvc\ such that
$x_{\mathrm{ZVC}}/R_c = 5$, where $x_{\mathrm{ZVC}}$ corresponds to
the intersection of the \zvc\ with the $x$-axis.

From eq.~(\ref{eq:rho}), we then compute the density map of each orbit
from a spectral analysis set by $F=10^{-3}$, giving position
expansions with 11~terms in $x$ and 17~in $y$ for the box orbit, and
13~terms in $x$ and 12~in $y$ for the loop orbit (right panels of
Fig.~\ref{fig:boxloopdens}).

\begin{figure}
\epsfxsize=0.9\columnwidth
\centerline{\epsfbox{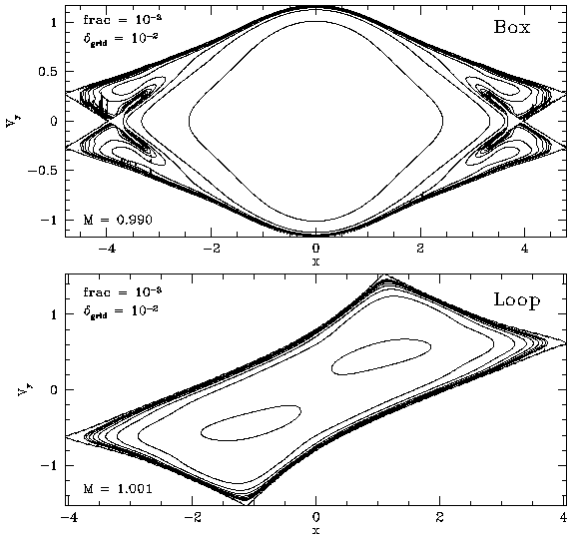}}
\caption{\emph{Upper panel}: \pvd{} for the box orbit, with
  contours ranging linearly from 0~to~$0.3$ in 11~steps.  Notice the
  symmetry of the \pvd{} with respect to $\dot{y}=0$.  \emph{Lower
    panel}: Same as previous, but for the loop orbit. The contours
  range linearly from 0~to~$0.2$ in 11~steps. Since the loop orbit has
  a definite sense of rotation (here, anti-clockwise), the \pvd{} only
  has reflection symmetry with respect to the point $(x, \dot{y}) =
  (0,0)$.}
\label{fig:vps}
\end{figure}

Fig.~\ref{fig:vps} presents the \pvd{} of the two previous orbits for
a~\los\ parallel to the $y$-axis.
The `wiggles' that can be seen in the middle of the loop orbit \pvd{}
(Fig.~\ref{fig:vps} lower panel) are a numerical artifact, at the
$\sim10\%$ level. The origin of this can easily be understood: for an
orbit in logarithmic potential (\ref{eq:logpot}), the characteristic
length scale, velocity and frequency are of the order
$x_{\mathrm{ZVC}}$, $v_0$ and $|\bomega| \sim v_0 / x_{\mathrm{ZVC}}$.
When we truncate the expansion (\ref{eq:xphi}) of the coordinate
$x(\bvarphi)$ according to the parameter $F$, we make a relative error
at the level ${X_\blambda / x_{\mathrm{ZVC}}} < F$ with the coordinate,
where $X_\blambda$ is the amplitude of the truncated term.  Meanwhile,
the corresponding term in the velocity expansion (\ref{eq:vphi}) has
an amplitude $\blambda\cdot\bomega \, X_\blambda$, where the frequency
of the expansion term $\blambda\cdot\bomega$ increases linearly as we
go to higher and higher order terms.  So the corresponding error with
the velocity is at the level ${\blambda\cdot\bomega \, X_\blambda / v_0}
\sim {|\blambda| X_\blambda / x_{\mathrm{ZVC}}} < |\blambda| F$.  So
while the same truncation reproduces the orbital position and
position-related quantities such as the density within a few percents,
it may significantly over- or under-estimate the high frequency
variations in the orbital velocity and for velocity-related quantities
such as \losvd's. A way to correct for this effect would be to
truncate the different expansions according to the higher amplitude
$X_\blambda$ or $\blambda\cdot\bomega \, X_\blambda$ at a given order
(Papaphilippou \& Laskar 1996). While truncating according to the
amplitudes $X_\blambda$ is straightforward, truncating according to
$\blambda\cdot\bomega \, X_\blambda$ requires a priori
knowledge of the BF decomposition $(\blambda,\bomega)$, which
generally would mean a double pass, and hence a significantly
increased number of computations. We have not followed this route
here, as this additional accuracy will play only a minor role in most
practical applications, such as the computation of the moments of the
\losvd.

\subsubsection{Surface of section and action space}

We have integrated a complete library of orbits in the logarithmic
potential (\ref{eq:logpot}). All the orbits have the same energy
$E=1.629$, and are integrated over typically $\sim 150$ periods with
4096~time-steps. The 140~initial conditions (ICs) are distributed
according to: 40~ICs evenly spaced in angle on the \zvc, and thus with
zero initial velocity so that they are boxes or irregular orbits, and
100~ICs evenly spaced in radius along the short axis ($x=0$ and
$y>0$), with an initial velocity vector perpendicular to the axis
($v_x>0$ and $v_y=0$). Those with $y < y_b$, where $y_b$ is the
amplitude of the last stable $y$-axis oscillation, are boxes or
irregular orbits. The remainder are loops. With the exception of the
closed loop at the chosen energy, these all cross the $y$-axis
perpendicularly in two points. We keep only those that cross at
apocenter, and furthermore remove all high-order resonant orbits,
leaving a total of 60 regular, open, `low-resonance' boxes and loops.
The spectral analysis of each integrated orbit is carried out with
$F=10^{-3}$, giving typically $\sim 13$ terms in the expansions along
$x$ and $y$.

\begin{figure}
\epsfxsize=0.9\columnwidth
\centerline{\epsfbox{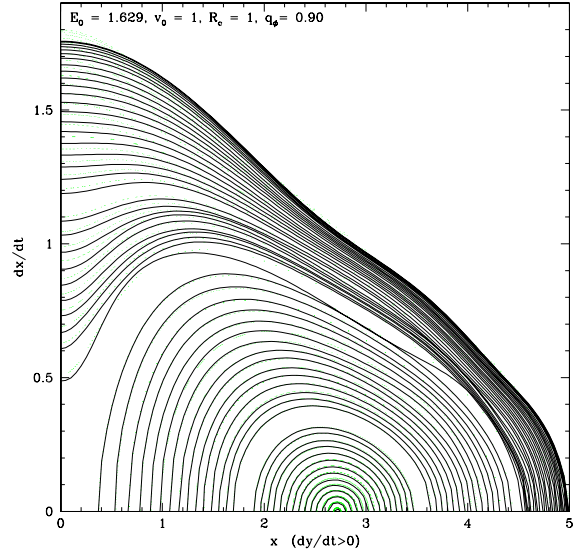}}
\caption{Surface of section $(x,\dot{x}),\;\dot{y}\geq 0$ computed from 
60~regular open orbits of energy $E=1.629$, for a truncation fixed by
$F=10^{-3}$. Only the low-resonance orbits which do not overlap over
themselves, have been kept. The empty areas correspond either to
stochastic areas or `high'-resonance islands. \emph{Dots} correspond
to the construction of the surface of section of each orbit during its
short numerical integration (\emph{see text}).}
\label{fig:surfofsec}
\end{figure}

We constructed the surface of section $(x,\dot{x}),\;\dot{y}\geq 0$
for this library by solving for the velocity $(\dot{x},\dot{y})$ at
each point $(x,y=0)$ of an orbit. The result is shown in
Fig.~\ref{fig:surfofsec}. Since we kept only low-resonance orbits,
this surface of section does not show any resonant island. However,
the discarded orbits leave empty areas in the surface of section
between the regular loops and boxes.  For comparison, we also show in
in Fig.~\ref{fig:surfofsec} the surface of section from the short
orbital integration integration ($\sim 150$ periods). As already noted
in \S\ref{sec:dmvp}, the velocity is less accurately retrieved in the
spectral process than the position, which results in some small
differences (few percents) in the surface of section. 

\begin{figure}
\epsfxsize=0.9\columnwidth
\centerline{\epsfbox{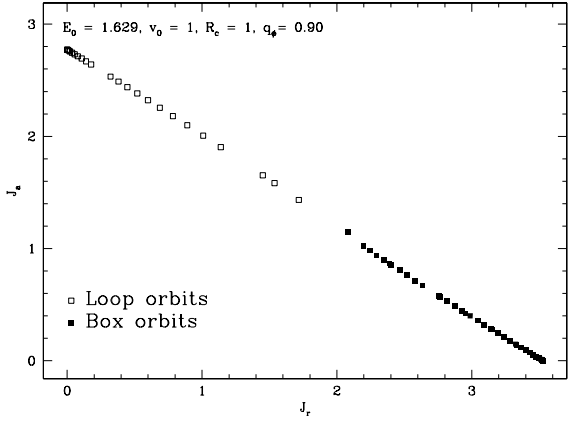}}
\caption{Action space ${\mathcal J}_a$, vs. ${\mathcal J}_r$ 
  computed from the same orbital set as Fig.~\ref{fig:surfofsec} and
  with the same truncation of the expansion. Note the natural
  distinction occurring between box orbits (\emph{solid squares}) and
  the loop orbits (\emph{open squares}). The two straight sequences
  meet in the middle as in Fig.~3-25 of BT87.  The gaps in the surface
  of section plot are seen here as well.  }
\label{fig:action}
\end{figure}

Using expression (\ref{eq:action}), we computed the actions associated
with an orbit, and constructed the action-space $({\mathcal
  J}_r,{\mathcal J}_a)$ of the mono-energetic library
(Fig.~\ref{fig:action}). This visualization of the library in
action-space is of particular interest: since the action-angle
variables are canonical, and their integration is particularly simple
(\S\ref{sec:action}), equal volume in action space is directly
associated to equal volume in phase-space (Binney \& Spergel 1984).
This provides a two-dimensional representation of the four-dimensional
phase-space. As expected, a natural dichotomy between box orbits
(solid squares) and loop orbits (open squares) appears in this space.

\subsection{Numerical discussion}
\label{sec:CPU}

The traditional way of computing the orbital density, just as any
other dynamical quantity such as the \losvd, is the following: along
the orbital motion (during the numerical integration), and for a
predetermined grid in configuration space, one counts up the number of
integration points falling in a given cell (e.g., Rix et al.\ 1997).
This gives the total fraction of time spent in this cell, which is
proportional to the local density after a long enough integration
time.

For a sufficiently fine grid, this process of `dropping balls in
buckets' can be considered as a random process, and the statistical
noise associated with a count of $N$ points in a cell is $\sim
\sqrt{N}$, leading to a precision in density of order $1/\sqrt{N}$.
The orbit in Fig.~\ref{fig:compare} was integrated for 4096~points,
and the density represented on a grid with step
$\delta_{\mathrm{grid}} = 10^{-2}$. Assuming that the integration
points are approximately uniformly distributed over the whole area
accessed by the orbit, roughly $2\times 0.5=1$ or
$1/\delta_{\mathrm{grid}}^2=10^4$ cells, the average number of points
per cell is $\sim 0.5$, leading to a very low precision in density ($>
100\%$). A precision of the same order of the one obtained with
spectral dynamics ($\la 5\%$), would require integrating $\sim 800$
times longer, in order to get $\sim 400$ points per cell.

\begin{figure}
\epsfxsize=0.9\columnwidth
\centerline{\epsfbox{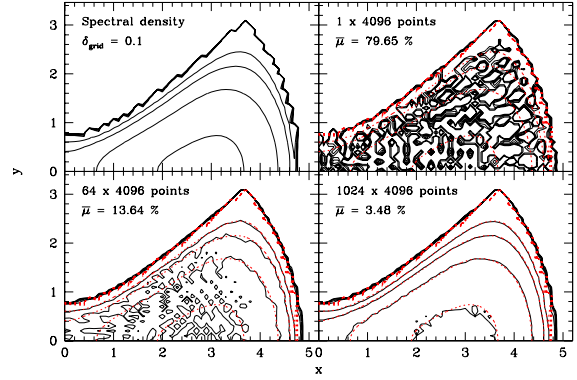}}
\caption{Comparison, for the $x>0,\,y>0$ part of the box orbit 
  of Fig.~\ref{fig:boxloopdens}, between the spectral density
  (\emph{upper left} panel and \emph{dotted} line in other panels) and
  the `bucket' densities (\emph{solid} line in other panels) for
  different integration times, related to the number of points. In
  both cases, contour levels range logarithmically from
  $1.35\,10^{-4}$ to $4\,10^{-4}$ in 4 steps, and the step of the grid
  used for the computation of the density is $\delta_{\mathrm{grid}} =
  0.1$.  The mean accuracy $\bar{\mu}$ is defined in the text. }
\label{fig:boxinteg}
\end{figure}

\begin{figure}
\epsfxsize=0.9\columnwidth
\centerline{\epsfbox{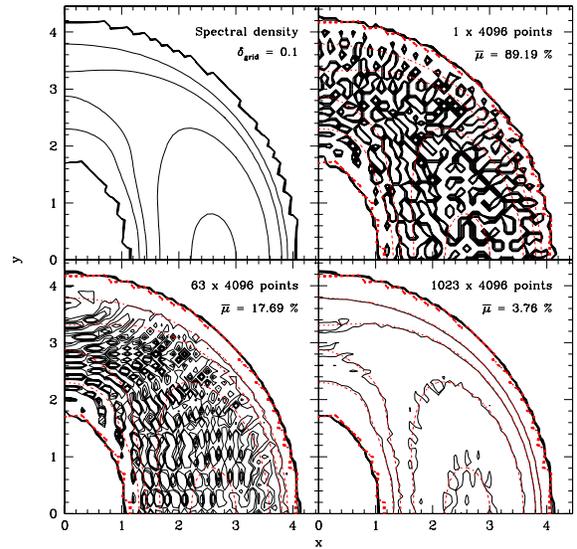}}
\caption{Same as Fig.~\ref{fig:boxinteg}, but for loop orbit of
  Fig.~\ref{fig:boxloopdens}. Contour levels range logarithmically
  from $1.1\,10^{-4}$ to $2.3\,10^{-4}$ in 4 steps.}
\label{fig:loopinteg}
\end{figure}

Fig.~\ref{fig:boxinteg} and \ref{fig:loopinteg} compare in a more
quantitative way the spectral density (\emph{dotted} line) and the
'bucket' densities (\emph{solid} line) computed for different
integration times for the box and loop orbits shown in
Fig.~\ref{fig:boxloopdens}. The estimated mean
accuracy $\bar{\mu}$ is defined as $\bar{\mu} \cor \langle
1/\sqrt{N_i} \rangle$, where $N_i$ is the number of balls in bucket
$i$, and $\langle\cdot\rangle$ is the average over all the non-empty
cells. As expected, the mean accuracy is directly related to the
number of integration points used during the computation of the
density.  It can be clearly seen in both cases that 4096~points along
the orbit, typically needed for the torus reconstruction from spectral
analysis, are \emph{highly} insufficient for computing a proper 
`bucket' density.

\begin{figure}
\epsfxsize=0.9\columnwidth
\centerline{\epsfbox{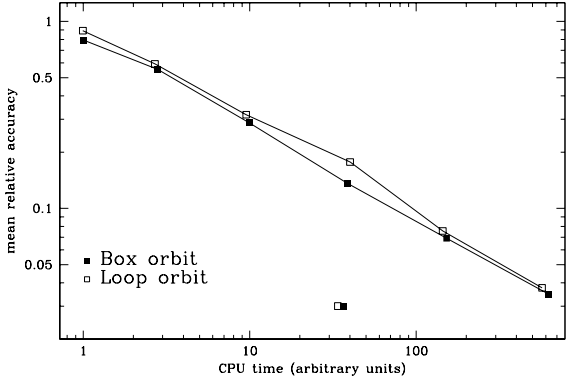}}
\caption{Mean accuracy of the density computations \emph{vs.}
  computing power needed for the computation, for the previous box
  (\emph{solid} symbols) and loop (\emph{open} symbols) orbits. Each
  dot on the lines corresponds to a case in Fig.~\ref{fig:boxinteg} or
  \ref{fig:loopinteg}, while the isolated symbols give an estimate of
  the spectral density mean accuracy ($\sim 3\%$).}
\label{fig:accuracy}
\end{figure}

Fig.~\ref{fig:accuracy} compares the different mean accuracies
obtained in terms of computing power. The CPU-time is expressed in
arbitrary units. Each density was computed on the same computer, and
the time used by each process was renormalized to the shortest one (a
few seconds on a standard Linux station). The spectral mean accuracy
is estimated to be $\sim 3\%$ (cf. \S\ref{sec:densmaps}). We conclude
that the spectral density is almost one order of magnitude more
accurate than the `bucket' density for the same computing power.

\section{Axisymmetric potentials}
\label{sec:3Daxi}

Recent work on modeling the internal dynamics of elliptical galaxies,
aimed at measuring masses of central black holes or the distribution
of extended dark matter, compare stellar absorption line kinematics
with fully general three-integral multi-component axisymmetric models.
While in some cases semi-analytic approaches can be used (e.g.,
Matthias \& Gerhard 1999), many studies use a variant of
Schwarzschild's orbit superposition technique (e.g., van der Marel et
al.\ 1998; Cretton \& van den Bosch 1999; Gebhardt et al.\ 1999;
Cretton, Rix \& de Zeeuw 1999). In this approach, orbits are
calculated numerically in a chosen potential, and their properties are
projected onto the plane of the sky.  These are then combined to
reproduce the observed surface density and \losvd's. The current
implementations use the `bucket' method for calculating the orbital
properties.

Motion in a three-dimensional axisymmetric potential can be reduced to
a two-dimensional problem by exploiting the conservation of the
$z$-component of the angular momentum $L_z$ (e.g., BT87), and
describing the motion in the meridional plane $(R,z)$ (the angular
variable $\phi$ being cyclic).  The properties of a regular orbit in
such a potential can thus be computed with much improved accuracy by
means of a minor adaptation of the spectral dynamics formalism
described in the above, and we present the relevant equations here,
together with some examples.  

The illustrations are made for a test-orbit with energy $E_0 = -0.8$
and angular momentum $L_z = 0.2$ integrated in the meridional plane of
a three-dimensional axisymmetric coreless logarithmic potential:
\beq
\Phi(R,z) \cor \frac{1}{2}v_0^2\, 
          \ln\left(R^2 + \frac{z^2}{q^2}\right),
\eeq
with $q = 0.9$ and $v_0 = 1$. Hereafter, we use $\alpha \cor i - \pi/2$,
where $i$ is the usual inclination angle of a galaxy ($i=0$ for a face-on
galaxy).


\subsection{Spatial density}

If the density of an orbit in the meridional plane is $\rho_M(R,z)$, the
associated spatial density is simply
\beq
\rho(R,z,\phi) = \frac{\rho_M(R,z)}{2\pi R},
\eeq
where the factor $2\pi$ ensures the normalization, so that $\int
\rho_M\,\d R\,\d z = \int \rho\,R\,\d R\,\d z\,\d\phi = 1$. 

\begin{figure}
\epsfxsize=0.9\columnwidth
\centerline{\epsfbox{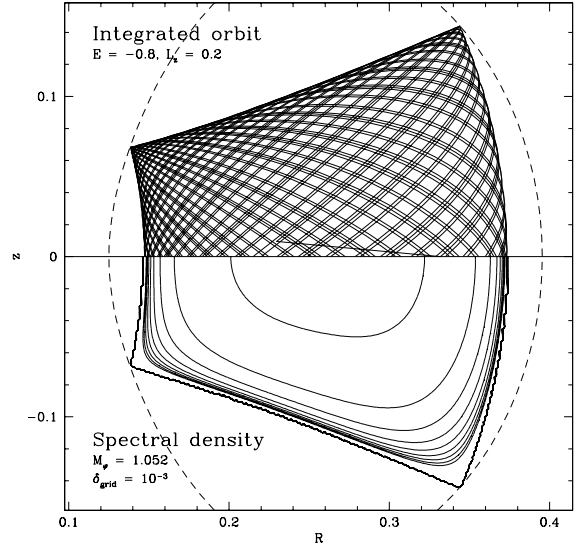}}
\caption{Three-dimensional axisymmetric orbit integrated in the 
  meridional plane (\emph{upper part}) and its associated meridional
  density computed from spectral analysis (\emph{lower part}).}
\label{fig:3Daxi-orbit}
\end{figure}

Fig.~\ref{fig:3Daxi-orbit} shows an example of the density of an orbit
in the meridional plane, reconstructed with the spectral formalism.

\subsection{Projected intensity}
\label{sec:3Dint}

\begin{figure}
\epsfxsize=0.9\columnwidth
\centerline{\epsfbox{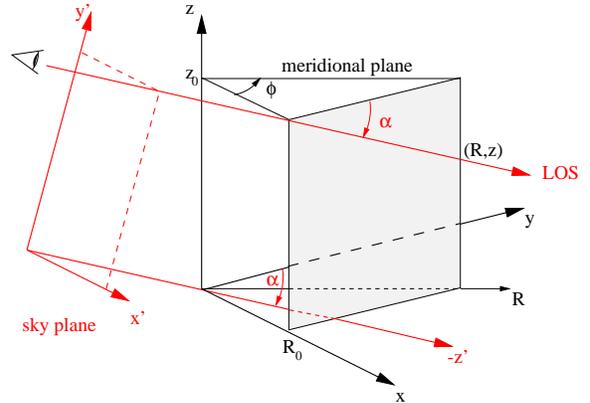}}
\caption{Notations used in the text for the conversion of two-dimensional 
to three-dimensional densities. We use $\alpha = i - \pi/2$, where $i$
is the customary inclination of the galaxy ($>0$ when seen from
above). }
\label{fig:3Daxi-geo}
\end{figure}

Cartesian coordinates $(x', y')$ on the plane of the sky relate to
spatial cylindrical coordinates $(R, \phi, z)$ as follows (see
Fig.~\ref{fig:3Daxi-geo}):
\beq
\left\{ \begin{array}{rcl}
        x' & = & R\cos\phi, \\
        y' & = & z\cos\alpha - R\sin\phi\sin\alpha, 
        \end{array}\right.
\eeq
where $\alpha$ defines the orientation of the line of sight. 
Assuming $\alpha \not = \pm\pi/2$ (i.e.,
the galaxy is not seen face-on), we find
\beq
\left\{ \begin{array}{rcl}
        R_{x',y'}(\phi) & = & |x'| / \cos\phi, \\ 
        z_{x',y'}(\phi) & = & y' / \cos \alpha + |x'|\tan\alpha \tan\phi.
        \end{array}\right.
\eeq
Furthermore, since $\d z' = -\d y/\cos \alpha$ and $\d y = R\d\phi / \cos
\phi$, 
\beq
\d z' = \frac{|x'| \d\phi}{\cos \alpha \cos^2\phi}.
\eeq
We can then compute the intensity on the sky $I_\alpha(x',y')\cor
\int_{\mathrm{los}}\rho(R,z,\phi)\,\d z'$ for $x' \not = 0$, to obtain: 
\beq
\label{eq:I3D}
I_\alpha(x',y') = \frac{1}{2\pi}\int_{-\pi/2}^{\pi/2}
\frac{\rho_M(R_{x',y'}(\phi),z_{x',y'}(\phi))\,\d\phi}{\cos\alpha \cos\phi}.
\eeq
The integrand is not singular at the limits of integration, since an
orbit reaches a maximum radius $R_{\mathrm{max}}$ in the meridional
plane, corresponding to a maximum angle $\phi_{\mathrm{max}} \cor
\arccos (R_0/R_{\mathrm{max}})$ (for $R_0 \leq R_{\mathrm{max}}$;
otherwise, we know that $I\equiv 0$).  Equivalently, the orbit reaches
a minimum radius $R_{\mathrm{min}}$, corresponding to the angle
$\phi_{\mathrm{min}} \cor \arccos (R_0/R_{\mathrm{min}})$ (or $0$ if
$R_0 > R_{\mathrm{min}}$). We can therefore restrict the integration
domain to the segments $[-\phi_{\mathrm{max}},-\phi_{\mathrm{min}}]
\cup [\phi_{\mathrm{min}},\phi_{\mathrm{max}}]$.

Similarly, for $x' = 0$, we obtain
\beq
\label{eq:I3DR0}
I_\alpha(x'=0,y') = \int_{\frac{R_{\mathrm{min}}}{\cos\alpha}}
        ^{\frac{R_{\mathrm{max}}}{\cos\alpha}} 
        \sum_\pm \rho_{y'}^\pm(z') \frac{\d z'}{z'\cos\alpha}.
\eeq
where we have defined $\rho_{y'}^\pm(z') \cor \rho_M(z'\cos\alpha,
\frac{y'}{\cos\alpha} \pm z'\sin\alpha)$.

\begin{figure}
\epsfxsize=0.9\columnwidth
\centerline{\epsfbox{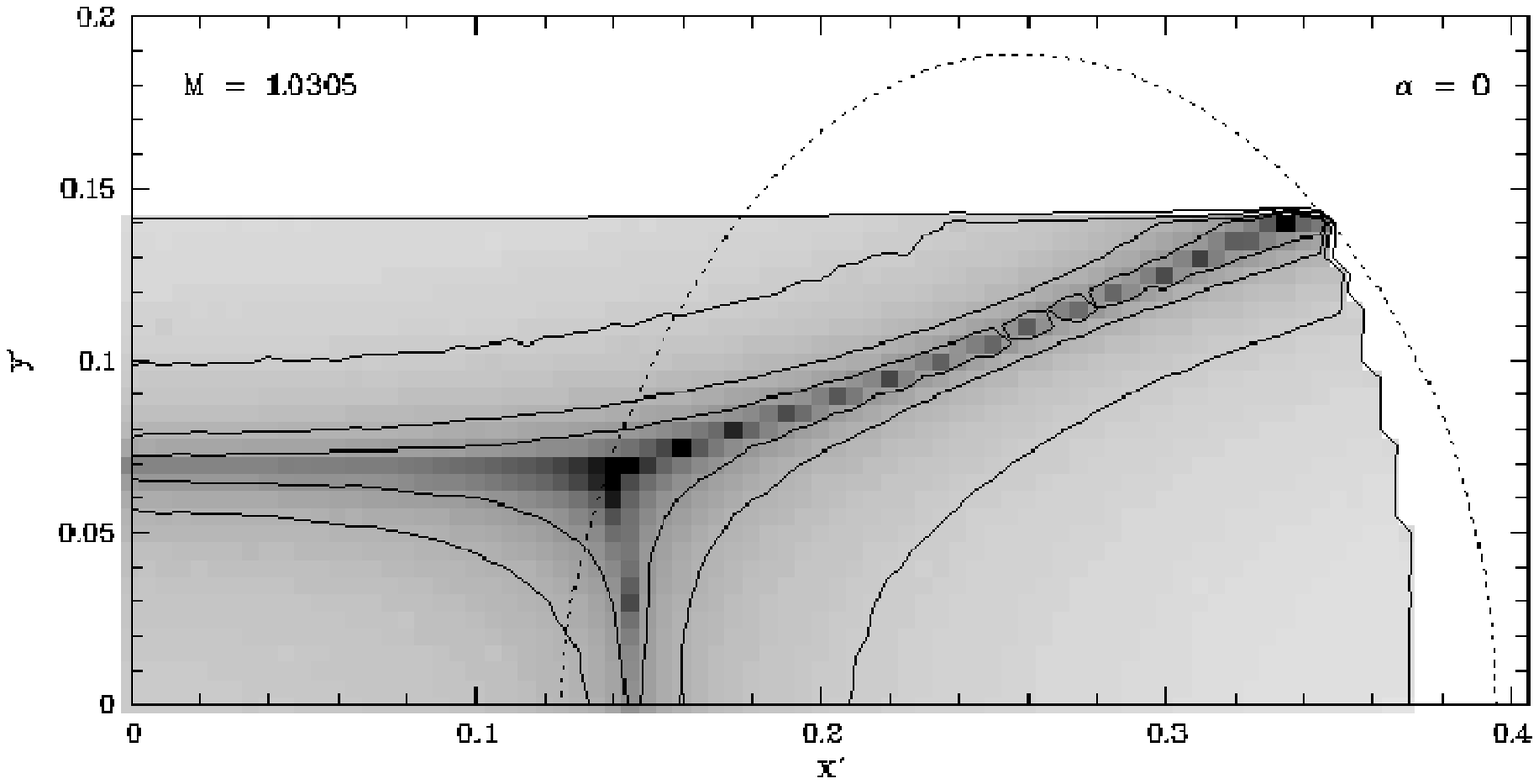}}
\epsfxsize=0.9\columnwidth
\centerline{\epsfbox{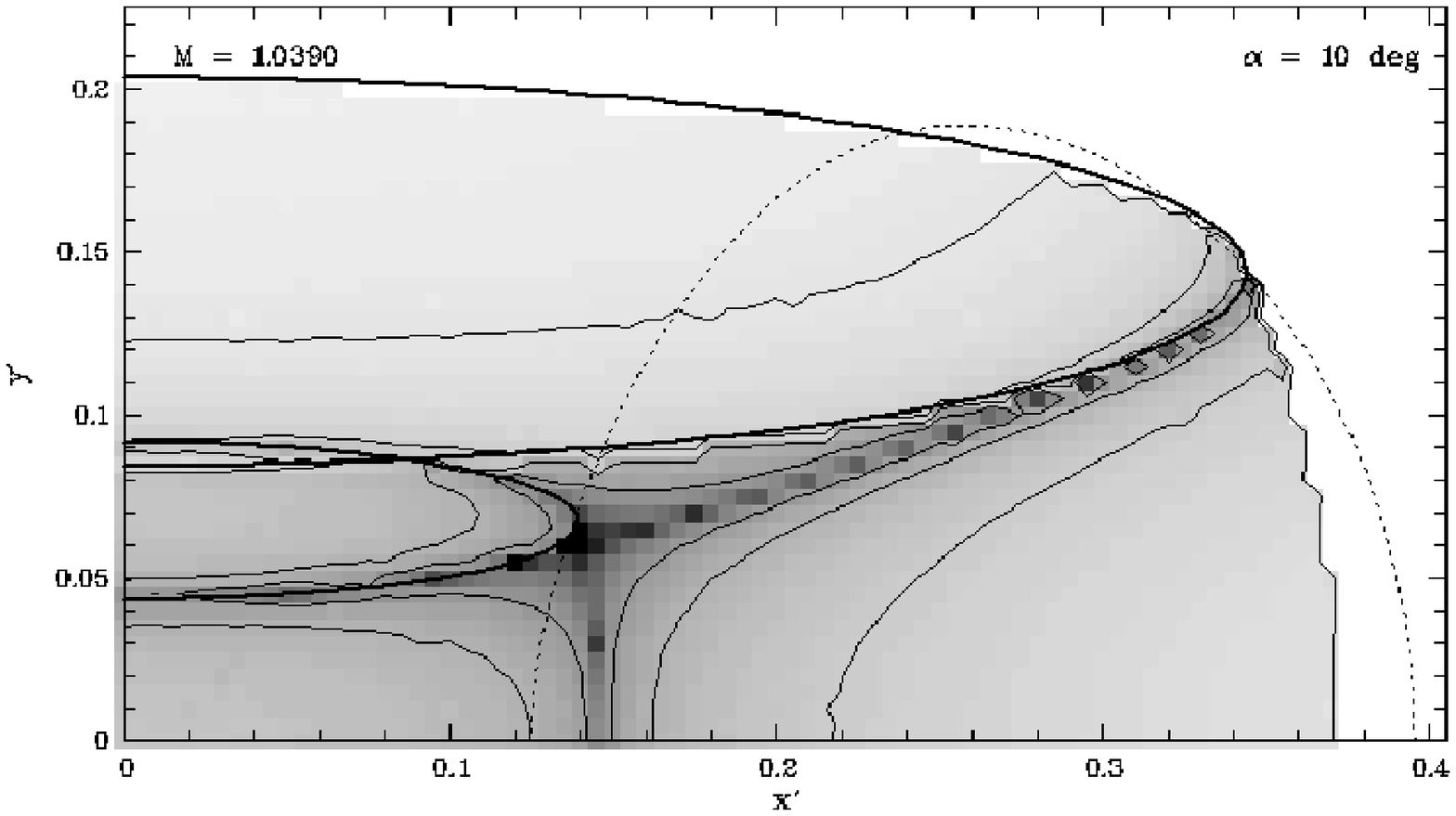}}
\caption{Sky density computed for the test-orbit seen edge-on
  ($\alpha=0$, \emph{upper panel}) and slightly overhead
  ($\alpha=10^\circ$, \emph{lower panel}). \emph{Dotted line}: \zvc{}
  in the meridional plane.  Note how the divergence of the density at
  the boundaries of the orbit in the meridional plane translates in
  sky density \emph{Heavy line}: geometrical location of the `corners'
  of the orbit.  The numerical line-of-sight integration precision is
  set to $10^{-2}$.}
\label{fig:3Daxi-dens}
\end{figure}

Fig.~\ref{fig:3Daxi-dens} shows the upper-right quadrant of the
projected density on the plane of the sky of the test orbit when seen
edge-on ($\alpha=0$) or slightly overhead ($\alpha=10^\circ$). As the
computation of the projected intensity involves an extra numerical
integration, the result is more sensitive to numerical errors in the
determination of the orbital density in the meridional plane.
Fig.~\ref{fig:3Daxi-dens} has been obtained with an integration
relative precision set to $10^{-2}$.


\subsection{Line-of-sight velocity distribution}
\label{sec:3Dlosvd}

\begin{figure}
\epsfxsize=0.9\columnwidth
\centerline{\epsfbox{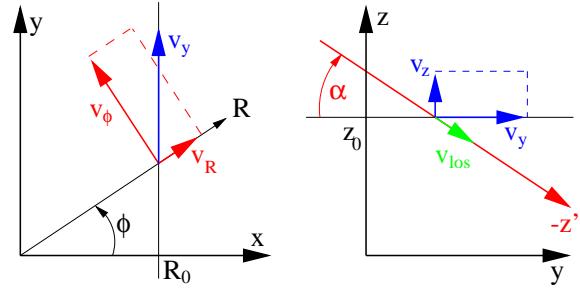}}
\caption{Elements of geometry for the computation of the velocity
along the \los.}
\label{fig:3Daxi-geovel}
\end{figure}

The velocity $v_{\mathrm{los}}$ along the \los\ 
(cf.\ Fig.~\ref{fig:3Daxi-geovel}) is given by
\beq
v_{\mathrm{los}} = \left(\frac{L_z}{R}\cos\phi + 
                   v_R \sin\phi\right) \cos\alpha + v_z \sin\alpha,
\eeq
where $v_R$ and $v_z$ are the velocity components in the meridional
plane, and $v_\phi = L_z/R$.  The \losvd{} can then be computed as
follows:
\beq
\label{eq:V3D}
{\mathrm{VP}}_{x',y'}(v_{\mathrm{los}}) = 
\frac{1}{(2\pi)^3}\sum_{i=1}^{M}\frac{1}{J^{(i)}},
\eeq
where the Jacobian
\beq
J^{(i)} \cor \frac{\partial(x',y',v_{\mathrm{los}})}
{\partial(\phi_1^{(i)},\phi_2^{(i)},\phi_3^{(i)})}, 
\eeq
is to be evaluated at the $i$-th solution $\varphi^{(i)} \equiv
(\phi_1^{(i)},\phi_2^{(i)})$, $i=1,\ldots,M$, of the system of
equations:
\beq
\left\{ \begin{array}{rcl}
    y' & = & z(\varphi^{(i)})\cos\alpha - 
      R(\varphi^{(i)})\sin\phi\sin\alpha, \\
    v_{\mathrm{los}} & = & \left(\frac{L_z}{R(\varphi^{(i)})}\cos\phi + 
      v_R(\varphi^{(i)}) \sin\phi\right) \cos\alpha \\
    & & \qquad\qquad\qquad\qquad\qquad +\; v_z(\varphi^{(i)}) \sin\alpha.
        \end{array}\right.
\eeq
The reason that we need only to solve for two action angles
$\phi_1^{(i)}$ and $\phi_2^{(i)}$ for the motion in the meridional
plane is that the third angle $\phi_3^{(i)}$ is simply the azimuthal
angle $\phi$, which is related to the two action angles
$\varphi^{(i)}$ by 
\beq
\phi_3^{(i)} \equiv \phi = \arccos\left[\frac{x'}{R(\varphi^{(i)})}\right].
\eeq

\begin{figure}
\epsfxsize=0.9\columnwidth
\centerline{\epsfbox{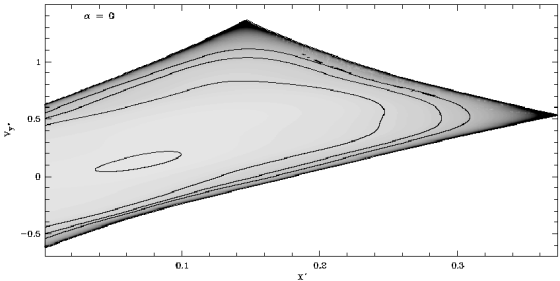}}
\caption{\pvd{} for a `virtual' long-slit placed along the $x'$-axis
  ($y'=0$) of the test-orbit seen edge-on ($\alpha=0$). The intensity
  at a point $(x',y'=0,v_{\mathrm{los}})$ is computed by means
  of eq.~\ref{eq:V3D}.}
\label{fig:3Daxi-vel}
\end{figure}

Fig.~\ref{fig:3Daxi-vel} shows the position-velocity diagram of the
test-orbit as seen with a `virtual' long-slit placed along the
$x'$-axis (that is $y'=0$). By symmetry, we can restrain the study to
the part $x'\geq 0$, where the velocities are expected to be mostly
positive since $L_z > 0$.

\section{Conclusions}
\label{sec:concl}

We have used the spectral analysis code provided by Carpintero \&
Aguilar (1997) to determine the properties of orbits in elliptic disks
and in axisymmetric potentials based on numerical integration over a
modest number of orbital periods. We have approximated regular orbits
by a truncated Fourier time series of a few tens of terms per
coordinate, and have reconstructed the underlying invariant torus by
computation of the associated action-angle variables. We have used
this to relate the uniform distribution of a regular orbit on its
torus to the non-uniform distribution in the familiar space of
observables by a simple Jacobian transformation between the two sets
of coordinates. This approach is by no means new, but we have extended
the published formalism to include calculation of the quantities used
in the construction of dynamical models by Schwarzschild's method, in
particular the orbital density and the observed \losvd's for elliptic
disks and for axisymmetric models.

In the standard implementation of Schwarzschild's method, orbital
properties are recorded on a grid of cells in configuration space $(x,
y, z)$ and in the space of observables $(x', y', v_{\mathrm{los}})$ (e.g.,
Cretton et al.\ 1999). The number of cells is generally taken to be of
the order of 1000-2000, and orbital densities and kinematic properties
are calculated by numerically integrating an orbit for a very long
time, so that each cell is crossed at least 100 times. Averaging of
the time spent in the cell, or of the velocity distribution in the
cell, then results in a $\sim 10\%$ accuracy on the orbital density and
the observed kinematics. By contrast, the spectral method allows the
invariant orbital torus to be described by a few tens of terms, and
allows computation of all the physical quantities in a
cell-independent way with much increased accuracy. At a given
accuracy, the spectral method requires less cpu time than the
traditional approach, at least for the two-dimensional cases we have
investigated. The resulting flexibility in the determination of the
orbital properties, and the drastic reduction of storage space for the
orbit library, provide further significant improvements in the
practical application of Schwarzschild's method.

The algorithm devised by CA98 uses only the orbital position for
spectral analysis. It truncates the orbital Fourier series according
to a criterion on the spectra of the coordinates, so that the
resulting Fourier series approximation and torus reconstruction is
less accurate in velocity space. Laskar's (1993) \textsc{naff} algorithm
uses the complete phase-space position, and may therefore be more
efficient in the determination of the base frequency of an orbit and
in the torus reconstruction. However, the precision obtained with CA98
seems adequate for the practical goals of the method.

The method employed here to compute orbital properties works best for
the main families of regular orbits. The higher-order resonances
(e.g., Miralda--Escud\'e \& Schwarzschild 1989) can be found by the
CA98 algorithm, but the torus reconstruction becomes more difficult
(e.g., Kaasalainen 1995). This suggests that the main advantage will
lie in the construction of dynamical models that have a fairly regular
phase space, such as models with central density profiles that are
only moderately cusped.

Irregular orbits cannot be represented by our Fourier
series expansions.  In principle it is possible to obtain the
properties of these orbits by subtracting the contributions of the
regular ones from an $f(E)$ or $f(E, L_z)$ component (e.g., Zhao 1996;
Cretton et al.\ 1999), which can be represented as the sum of all
regular and irregular orbits at that $E$ (and $L_z$). Use of the
smooth regular orbits and the $f(E)$ and $f(E,L_z)$ components as
building blocks in Schwarzschild's method therefore means that the
irregular orbits are included as well.

Application of spectral dynamics to compute the observable properties
of regular orbits in genuinely three-dimensional potentials is
possible in principle. Construction of triaxial dynamical models that
fit observed the kinematics of early-type galaxies may well benefit
from the same approach, although the rich phase space, and the
presence of many irregular orbits, may complicate its practical
application (e.g., H\"afner et al.\ 1999).

\medskip
It is a pleasure to thank Luis Aguilar and Daniel Carpintero for
making available their spectral dynamics code, and to thank Wyn Evans
for constructive comments.


\appendix


\section{Sridhar \& Touma potential}
\label{app:stackel}

The test-orbit used in \S\ref{sec:numstack} was integrated in the
two-dimensional separable potential introduced by Sridhar \& Touma (1997,
ST). We collect here a number of properties of the orbits in this
potential.


The ST potential is defined by
\beq
\label{eq:ST97}
\Phi(r,\theta) \cor r^\alpha
        \left[(1+\cos\theta)^{(1+\alpha)} 
       + (1-\cos\theta)^{(1+\alpha)} \right].
\eeq
It is of St\"ackel form in parabolic coordinates $\xi = r(\cos\theta +
1) \geq 0$ and $\eta = r(\cos\theta - 1) \leq 0$, in which it takes
the form
\beq
\Phi = \frac{F_+(\xi)}{\xi-\eta} + \frac{F_-(\eta)}{\eta-\xi}, 
\eeq
with $F_+(\xi) \cor 2\xi^{1+\alpha}$ and $F_-(\eta) \cor
-2|\eta|^{1+\alpha}$.

In addition to the energy
\beq
E = \frac{2}{\xi-\eta} (\xi p_\xi^2 - \eta p_\eta^2) + \Phi,
\eeq 
every orbit has a second isolating integral of motion $I_2$, given by
\beq
I_2 = 2\xi\,p_\xi^2 - \xi E + F_+(\xi) = 2\eta\,p_\eta^2 - \eta E + F_-(\eta).
\eeq


As eq.~(\ref{eq:ST97}) shows, the ST potential is scale-free, so that
the shape of an orbit is determined only by the value of the second
integral, while its scale is related to the energy. Two orbits of
opposite $I_2$ are symmetric with respect to the $x$-axis; we consider
hereafter only the case $I_2 \geq 0$.

Defining the functions $I_+(\eta) \cor -\eta E + F_-(\eta)$ and
$I_-(\xi) \cor - \xi E + F_+(\xi)$ for a given energy $E$, the
boundaries of an orbit with second integral $I_2\;(\geq 0)$ are given
by the roots of the equations $I_+(\eta) = I_2$ (2 roots $\eta_1$ and
$\eta_2$) and $I_-(\xi) = I_2$ (one root $\xi_0$; see e.g.,
Fig.~\ref{fig:reconst}).


The orbital~\df\ of an orbit with energy $E_0$ and second integral
$I_0$ can be written (as a consequence of the strong Jeans theorem)
as:
\beq
f_{E_0,I_0}(E,I_2) \propto \delta(E-E_0)\,\delta(I_2-I_0).
\eeq
The orbital surface density is as usually given by
$\rho_{E_0,I_0}(r,\theta) \cor \int f_{E_0,I_0} \d^2v$.
We have $\d\xi\d\eta\d p_\xi\d p_\eta = \d\vecr\d\vecv$, which
translates to:
\beq
\d^2v = \frac{1}{r}\frac{\partial(\xi,\eta)}{\partial(r,\theta)}
        \frac{\partial(p_\xi,p_\eta)}{\partial(E,I_2)}\,\d E\d I_2, 
\eeq
with
\beq
\frac{\partial(\xi,\eta)}{\partial(r,\theta)} = 2r\,\sin\theta, 
\eeq
and 
\beq
\frac{\partial(p_\xi,p_\eta)}{\partial(E,I_2)} = 
        \frac{1}{2}\frac{1}{\sqrt{(I_+(\eta) - I_0)(I_0 - I_-(\xi))}}.
\eeq
Therefore, the unnormalized orbital density $\rho^\star$ is 
\beq
\rho^\star_{E_0,I_0}(\xi,\eta) = \frac{1}{2}
\frac{1}{\sqrt{(I_+(\eta) - I_0)(I_0 - I_-(\xi))}}, 
\eeq
while the total mass of this unnormalized orbit is found to be:
\bey
M^\star_{E_0,I_0} & \cor & \int \rho^\star_{E_0,I_0} \,\d^2\vecr \\
        & = & \frac{1}{4}\left[
\int_{\eta_1}^{\eta_2}\frac{\sqrt{\eta}}{\sqrt{\delta_\eta}}\d\eta
\int_0^{\xi_0}\frac{\d\xi}{\sqrt{\xi}\sqrt{\delta_\xi}}\right. \\
        & & \qquad\qquad + \left.
\int_{\eta_1}^{\eta_2}\frac{\d\eta}{\sqrt{\eta}\sqrt{\delta_\eta}}
\int_0^{\xi_0}\frac{\sqrt{\xi}}{\sqrt{\delta_\xi}}\d\xi \right],
\eey
with shorthand notations $\delta_\eta \cor I_+(\eta) - I_0$ and
$\delta_\xi \cor I_0 - I_-(\xi)$.  This double integration can be
carried out easily by numerical means.

\label{lastpage}

\end{document}